\documentclass[12pt,a4paper,final]{revtex4}

\usepackage{sidecap}
\usepackage{ulem}
\usepackage{epsfig}
\usepackage{amsmath,amssymb,amsthm}
\usepackage{graphicx}
\usepackage{bm}
\usepackage{color}

\DeclareMathAlphabet{\mathpzc}{OT1}{pzc}{m}{it}

\begin{document}

\renewcommand{\textfraction}{0.00}


\newcommand{\vAi}{{\cal A}_{i_1\cdots i_n}} 
\newcommand{\vAim}{{\cal A}_{i_1\cdots i_{n-1}}} 
\newcommand{\vAbi}{\bar{\cal A}^{i_1\cdots i_n}}
\newcommand{\vAbim}{\bar{\cal A}^{i_1\cdots i_{n-1}}}
\newcommand{\htS}{\hat{S}} 
\newcommand{\htR}{\hat{R}}
\newcommand{\htB}{\hat{B}} 
\newcommand{\htD}{\hat{D}}
\newcommand{\htV}{\hat{V}} 
\newcommand{\cT}{{\cal T}} 
\newcommand{\cM}{{\cal M}} 
\newcommand{\cMs}{{\cal M}^*}
\newcommand{\vk}{\vec{\mathbf{k}}}
\newcommand{\bk}{\bm{k}}
\newcommand{\kt}{\bm{k}_\perp}
\newcommand{\kp}{k_\perp}
\newcommand{\km}{k_\mathrm{max}}
\newcommand{\vl}{\vec{\mathbf{l}}}
\newcommand{\bl}{\bm{l}}
\newcommand{\bK}{\bm{K}} 
\newcommand{\bb}{\bm{b}} 
\newcommand{\qm}{q_\mathrm{max}}
\newcommand{\vp}{\vec{\mathbf{p}}}
\newcommand{\bp}{\bm{p}} 
\newcommand{\vq}{\vec{\mathbf{q}}}
\newcommand{\bq}{\bm{q}} 
\newcommand{\qt}{\bm{q}_\perp}
\newcommand{\qp}{q_\perp}
\newcommand{\bQ}{\bm{Q}}
\newcommand{\vx}{\vec{\mathbf{x}}}
\newcommand{\bx}{\bm{x}}
\newcommand{\tr}{{{\rm Tr\,}}} 
\newcommand{\bc}{\textcolor{blue}}

\newcommand{\beq}{\begin{equation}}
\newcommand{\eeq}[1]{\label{#1} \end{equation}} 
\newcommand{\ee}{\end{equation}}
\newcommand{\bea}{\begin{eqnarray}} 
\newcommand{\eea}{\end{eqnarray}}
\newcommand{\beqar}{\begin{eqnarray}} 
\newcommand{\eeqar}[1]{\label{#1}\end{eqnarray}}
 
\newcommand{\half}{{\textstyle\frac{1}{2}}} 
\newcommand{\ben}{\begin{enumerate}} 
\newcommand{\een}{\end{enumerate}}
\newcommand{\bit}{\begin{itemize}} 
\newcommand{\eit}{\end{itemize}}
\newcommand{\ec}{\end{center}}
\newcommand{\bra}[1]{\langle {#1}|}
\newcommand{\ket}[1]{|{#1}\rangle}
\newcommand{\norm}[2]{\langle{#1}|{#2}\rangle}
\newcommand{\brac}[3]{\langle{#1}|{#2}|{#3}\rangle} 
\newcommand{\hilb}{{\cal H}} 
\newcommand{\pleft}{\stackrel{\leftarrow}{\partial}}
\newcommand{\pright}{\stackrel{\rightarrow}{\partial}}

\title{Explaining the fine hierarchy in pion and kaon suppression at LHC:
Importance of fragmentation functions}

\date{\today}
 
\author{Magdalena Djordjevic}
\affiliation{Institute of Physics Belgrade, University of Belgrade, Serbia}
\author{Marko Djordjevic}
\affiliation{Faculty of Biology, University of Belgrade, Serbia}

\begin{abstract} 
We here concentrate on available $\pi^\pm$ and $K^\pm$ ALICE preliminary $R_{AA}$ data in central 2.76 TeV Pb+Pb collisions at LHC. These data show an interesting fine resolution hierarchy, i.e. the measured $K^\pm$ data have consistently lower suppression compared to $\pi^\pm$ measurements. We here ask whether theoretical predictions based on energy loss in dynamical QCD medium can quantitatively and qualitatively explain such fine resolution. While our suppression calculations agree well with the data, we find that qualitatively explaining the fine hierarchy  critically depends on the choice of fragmentation functions. While the most widely used fragmentation functions lead to the reversal of the observed hierarchy, a more recent version  correctly reproduce the experimental data. We here point to the reasons behind such discrepancy in the predictions. Our results argue that accuracy of the theoretical predictions reached a point where comparison with fine resolution data at LHC can generate useful understanding.
\end{abstract}

\pacs{12.38.Mh; 24.85.+p; 25.75.-q}

\maketitle 

\section{Introduction} 

Jet suppression~\cite{Bjorken} measurements in experiments involving ultra-relativistic heavy ion collisions - and their subsequent comparison with theoretical predictions - provide an excellent tool 
for studying the properties of a QCD medium created in these 
collisions~\cite{Gyulassy,DBLecture,Wiedemann2013,Vitev2010}. The suppression results from 
the energy loss of high energy partons moving through the 
plasma~\cite{suppression,BDMS,BSZ,KW:2004}; consequently, reliable computations 
of jet energy loss  are essential for the reliable predictions of 
jet suppression. While several theories of jet energy loss provide a reasonable agreement with specific measured data~\cite{Majumder_2012,Buzzatti,Zapp2013,Wang2011,MD_PRC2012}, there is a question to what extent theoretical predictions can explain fine resolution between different observables.
An interesting example of such fine resolution is provided by recently available ALICE preliminary measurements~\cite{ALICE_preliminary}  for charged pions and kaons in central 2.76 TeV  Pb+Pb collisions at LHC. These data show that $K^\pm$ suppression is systematically (across the entire momentum range) somewhat larger than $\pi^\pm$ suppression. Consequently, these data provide an interesting example of fine qualitative differences between observables, and  also an opportunity to test if theory can resolve such fine hierarchy. To answer this question, we here use our theoretical formalism for jet suppression in finite size dynamical QCD medium~\cite{MD_LHSupp_2013} to test whether, and under what conditions, this formalism is able to explain such measurements. 

\section{Numerical procedure}

The quenched hadron spectra $\frac{E_f d^3\sigma(H_Q)}{dp_f^3}$ is  calculated
from the initial spectra $\frac{E_i d^3\sigma(Q)}{dp^3_i}$ by using the generic pQCD convolution
\begin{eqnarray}
\frac{E_f \, d^3\sigma(H_Q)}{dp_f^3} &=& \;  \frac{E_i \, d^3\sigma(Q)}{dp^3_i} \nonumber \\
& \otimes&
{P(E_i \rightarrow E_f )}
\otimes D(Q \to H_Q), \;
\label{schem} \end{eqnarray}
where $Q$ denotes light quarks and gluons (we neglect heavy quarks since their contribution to pions and kaons is negligible). In the equation above $P(E_i \rightarrow E_f )$ denotes the energy loss 
probability, while $D(Q \to H_Q)$ represents the fragmentation function of quarks or gluons $Q$ to hadron $H_Q$.  

The energy loss probability $P(E_i \rightarrow E_f)$ 
is generalized to include both radiative and collisional energy loss in a 
realistic finite size dynamical QCD medium. In the calculation of the energy loss probability, we also included multi-gluon~\cite{GLV_suppress} and path-length fluctuations~\cite{WHDG}. In the path length fluctuations, the length distributions for 0-5\% most central collisions are introduced according to~\cite{Dainese}. Note that the path length distribution is a geometric quantity, which is the 
same for all jet varieties. 

Furthermore, we include the procedure for gluon number fluctuations in the radiative energy loss probability as described in detail in Ref.~\cite{MD_LHSupp_2013}. We recently improved the dynamical energy loss formalism in the  finite size QCD medium~\cite{MD_PRC,DH_PRL} to include finite magnetic mass effects~\cite{MD_MagnMass} and running coupling~\cite{MD_LHSupp_2013}. Specifically, Eq. (10) in~\cite{MD_MagnMass} represents the gluon radiation spectrum used in our calculations, while running coupling is introduced according to Eqs. (3) and (4) from~\cite{MD_LHSupp_2013}. The full fluctuation spectrum for collisional energy loss is taken to be Gaussian, whose mean is equal to the average collisional
energy loss, and the width is determined by $\sigma_{coll}^2 = 2 T \langle \Delta E^{coll}(p_\perp,L)\rangle 
$~\cite{Moore:2004tg}. Note that the average collisional energy loss $\Delta E^{coll}(p_\perp,L)$ is determined by 
Eq.~(14) in~\cite{MD_Coll}, $T$ denotes the temperature of the medium, $p_\perp$ is 
the initial jet momentum, while $L$ is the jet path length.

Finally, note that in the suppression calculations  we first calculate how the quark and gluon spectrums are modified by the radiative energy loss, and subsequently calculate how this spectrum changes due to collisional energy loss. That is, we separately treat radiative and collisional energy losses in the suppression calculations. Such approximation is reasonable when  the radiative and collisional energy losses are sufficiently small (which is in the essence of the soft-gluon, soft-rescattering
approximation, assumed in all available energy loss calculations), and when radiative and collisional energy losses can be decoupled (as is the 
case in the HTL approach~\cite{RadVSColl} that is used in our energy loss 
calculations~\cite{MD_PRC,DH_PRL,MD_Coll,MD_MagnMass}). 
We also assume a large enough quenched energy $E_f$, so that we can apply  
the Eikonal approximation. Finally, we assume that the jet to hadron fragmentation functions are the same in Pb+Pb and  
$e^+e^-$ collisions, which is a reasonable approximation for the deconfined 
QCD medium, i.e. when hadronization occurs 
after the parton leaves the Quark-Gluon Plasma (QGP).  

\section{Numerical results}

We consider a 
QGP of temperature $T{\,=\,}304$\,MeV (as extracted by ALICE~\cite{Wilde2012}), with 
$n_f{\,=\,}2.5$ effective light quark flavors; we take perturbative QCD scale to be $\Lambda_{QCD}=0.2$~GeV. Mass of the light quarks is assumed to be dominated by the thermal mass $M{\,=\,}\mu_E/\sqrt{6}$, while the gluon mass is  $m_g=\mu_E/\sqrt{2}$~\cite{DG_TM}. Here Debye mass $\mu_E \approx 0.9$~GeV is obtained by self-consistently solving the  Eq.~(3) from~\cite{MD_LHSupp_2013} (see also~\cite{Peshier2006}), and magnetic mass $\mu_M$ is taken as $0.4 \, \mu_E < \mu_M < 0.6 \, \mu_E$~\cite{Maezawa,Bak}. Initial distributions for  gluons and light quarks are computed at next-to-leading order as  in~\cite{Vitev0912}, and path-length fluctuations are taken from~\cite{Dainese}. Since fragmentation functions can affect hierarchy of the suppression predictions for different particle species~\cite{Vitev06}, we will here in parallel use two choices, i.e. the widely used KKP~\cite{KKP} and a more recent DSS~\cite{DSS} fragmentation functions. 

\begin{figure*}
\epsfig{file=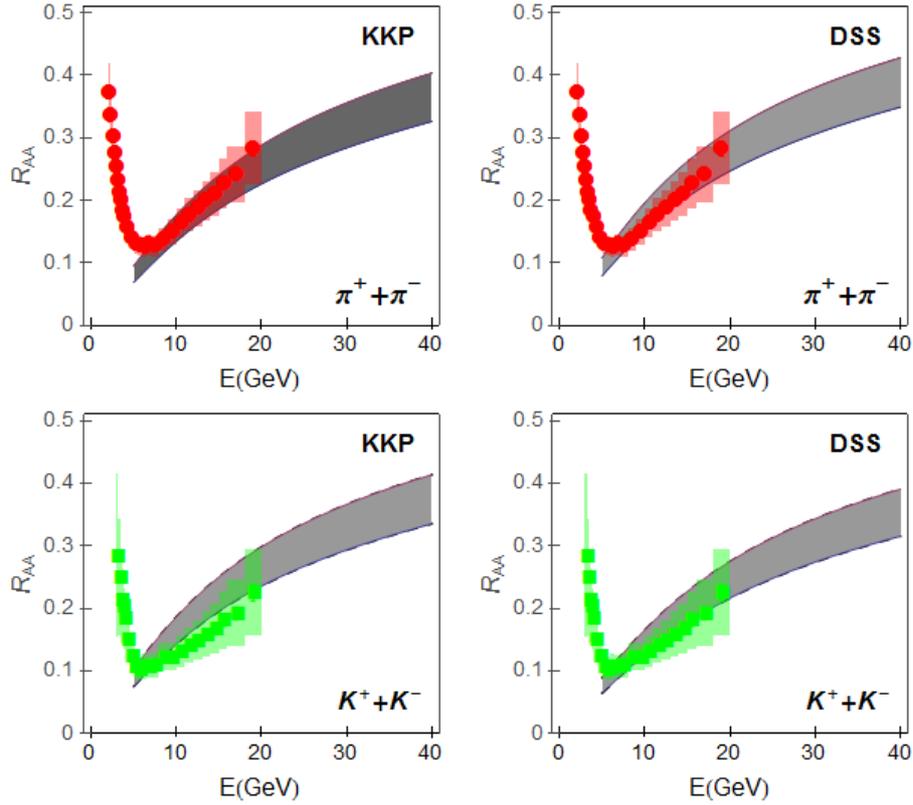,width=4.7in,height=4.2in,clip=5,angle=0}
\vspace*{-0.5cm}
\caption{{\bf Theory vs. experiment for momentum dependence of pion and kaon $R_{AA}$.} The upper panels show the comparison of $\pi^\pm$ 0-5\% central 2.76 TeV Pb+Pb ALICE preliminary~\cite{ALICE_preliminary} $R_{AA}$ data (red circles) with the pion suppression predictions, by using KKP~\cite{KKP} (the left panel) and DSS~\cite{DSS} (the right panel) fragmentation functions. The two lower panels show the analogous comparison for $K^\pm$ data. On each 
panel, the gray region corresponds to the case when $0.4 < \mu_M/\mu_E < 0.6$, where the upper (lower) 
boundary of each band corresponds to $\mu_M/\mu_E =0.4$ ($\mu_M/\mu_E =0.6$).}
\label{PionKaonRAA}
\end{figure*}

Since the goal of this paper is to test our ability to explain the fine resolution between the pion and kaon suppression data, we start by individually comparing these data with our predictions. Figure~\ref{PionKaonRAA} shows comparison of the theoretical predictions with preliminary $\pi^\pm$ and $K^\pm$ data for KKP and DSS fragmentation functions. For pions, we see a very good agreement between our predictions and experimental data, irrespectively of the choice of the fragmentation functions. For kaons, we again notice a good agreement with measurements. However, one can also notice a better agreement for DSS compared to KKP fragmentation functions, i.e. KKP functions lead to consistently lower kaon suppression predictions compared to the data. Consequently, there is a question whether differences in the fragmentation functions can affect explanation of the observed fine hierarchy of the data.

\begin{figure*}
\epsfig{file=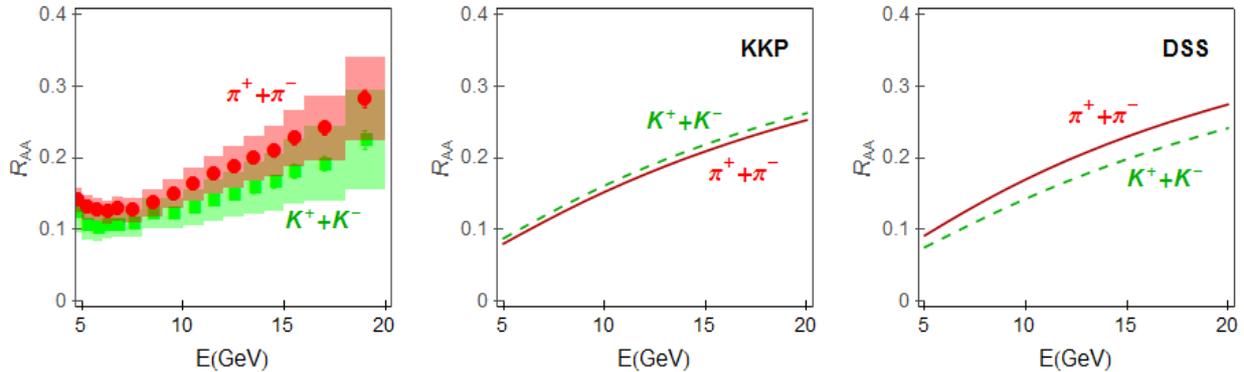,width=6.4in,height=2.1in,clip=5,angle=0}
\vspace*{-0.7cm}
\caption{{\bf Comparison of pion and kaon $R_{AA}$.} The left panel show the comparison between $\pi^\pm$ (red circles) and $K^\pm$ (green squares)  for 0-5\% central 2.76 TeV Pb+Pb ALICE preliminary~\cite{ALICE_preliminary} $R_{AA}$ data.  The central panel shows comparison between the theoretical predictions for pion and kaon suppression, by using KKP fragmentation functions. The full and the dashed curves respectively correspond to  $\pi^\pm$ and $K^\pm$ suppression predictions. A fixed magnetic to electric mass ratio $\mu_M/\mu_E =0.5$ is used. The right panel shows the analogous comparison for DSS fragmentation functions. }
\label{KPSupp}
\end{figure*}

This issue is further investigated in Figure~\ref{KPSupp}, where in the left panel we show the experimentally observed hierarchy in  $\pi^\pm$ and $K^\pm$ data. The central and the right panel in the Fig.~\ref{KPSupp} show, respectively,  predictions for this hierarchy by using KKP and DSS fragmentation functions. Note that, for clearer comparison, we fix magnetic to electric mass ratio to $\mu_M/\mu_E =0.5$.  The left panel (experimental measurements) shows that $K^\pm$ $R_{AA}$ data points are consistently bellow the corresponding  $\pi^\pm$ $R_{AA}$ data. However, theoretical calculations using KKP fragmentation functions show a reversed hierarchy, i.e. it is obtained that the predicted $K^\pm$ $R_{AA}$ is consistently above  the corresponding  $\pi^\pm$ $R_{AA}$. On the other hand, DSS fragmentation functions clearly reproduce the correct (experimentally observed) hierarchy both quantitatively and qualitatively. 

\begin{figure*}
\epsfig{file=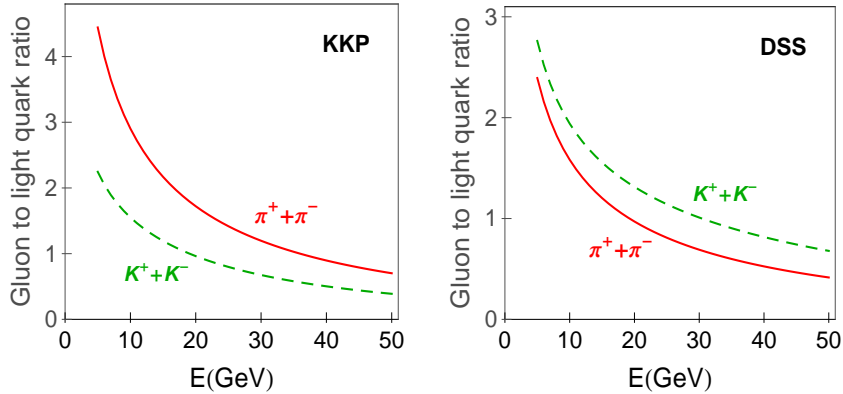,width=4.4in,height=2.2in,clip=5,angle=0}
\vspace*{-0.7cm}
\caption{{\bf Ratio of gluon to light quark contribution  in initial distributions of pions and kaons as a function of momentum. } The left and the right panel show the comparison of  gluon to light quark contribution ratio in the initial distributions of pions and kaons, by using KKP and DSS fragmentation functions, respectively. On each panel, the full curve corresponds to the pion case, while dashed curve corresponds to the kaon case.}
\label{GLRatio}
\end{figure*}

We will bellow investigate the reasons for such qualitative difference in predictions when different fragmentation functions are used. We start by noting that the gluon suppression is significantly larger compared to the light quark suppression, which is a consequence of a much larger energy loss for gluon jets compared to light quark jets. This difference in the suppressions rises a question of what is relative gluon to light quark contribution in pion and kaon distributions, when these two types of fragmentation functions are used. This ratio is shown in Figure~\ref{GLRatio}, where in the left panel, we see that KKP fragmentation functions predict a larger gluon to light quark ratio  for pions compared to kaons. This ratio evidently leads to the larger suppression of pions compared to kaons, having in mind the larger suppression in gluons compared to light quarks. On the other hand, DSS fragmentation functions lead to a larger gluon contribution in kaons compared to pions, which evidently leads to larger suppression of kaons compared to pions, as shown in the right panel of Fig.~\ref{KPSupp} - this time in agreement with the experimental data. Consequently, we see that the difference in the gluon to light quark contribution in the fragmentation functions leads to qualitative differences in the predicted suppression hierarchy.

\section{Conclusions} 

Comparison of jet suppression predictions with the available experimental data, is considered an excellent tool to test our understanding of QCD matter created in ultra relativistic heavy ion collisions. We here concentrated on recently available $\pi^\pm$ and $K^\pm$ $R_{AA}$ data in central Pb+Pb collisions at LHC, where we observed an interesting fine hierarchy between the measured suppression patterns. Such fine resolution allows more precisely testing to what extent, and under what numerical/computational conditions,  these experimental results can be theoretically explained. We particularly concentrated on the role of fragmentation functions in this study, since they have potential to affect hierarchy of the suppression predictions for different particle species. We found that our predictions lead to an excellent agreement with the measured data for DSS fragmentation functions, and somewhat worse - but still reasonable - agreement for KKP fragmentation functions.  However, we also found that qualitatively reproducing the experimentally observed fine hierarchy critically depends of the  choice of the fragmentation functions. This not only underscores an importance on proper choice of fragmentation functions, but also argues about usefulness of comparing the theoretical predictions with suppression data even at the fine resolution exemplified here. 

\bigskip

{\em Acknowledgments:} 
This work is supported by Marie Curie International Reintegration Grants 
within the $7^{th}$ EU Framework Programme 
(PIRG08-GA-2010-276913 and PIRG08-GA-2010-276996) and by the Ministry of Education, Science and Technological 
Development of the Republic of Serbia, under projects No. ON171004 and 
ON173052 and by L'OREAL-UNESCO National Fellowship in Serbia.  We thank I. Vitev and Z. Kang for providing the initial light 
flavor distributions and useful discussions. 
We also thank ALICE collaboration for providing the shown preliminary data, and M. 
Stratmann and Z. Kang for help with DSS fragmentation functions.

\end{document}